# Introduction of interactive learning into French university physics classrooms*

Alexander L. Rudolph,[1,2] Brahim Lamine,[3,4] Michael Joyce,[3] Hélène Vignolles,[3] and David Consiglio[5]

[1]*Department of Physics and Astronomy, California State Polytechnic University, Pomona, California 91768, USA; alrudolph@csupomona.edu*

[2]*Professeur Invité, Université Pierre et Marie Curie, Paris, France*

[3]*Faculté de Physique, Université Pierre et Marie Curie, Paris, France*

[4]*Faculté des Sciences et de l'Ingénieur, Université Paul Sabatier, Toulouse, France*

[5]*Bryn Mawr College, Bryn Mawr, Pennsylvania 19010, USA*

# Abstract

We report on a project to introduce interactive learning strategies (ILS) to physics classes at the *Université Pierre et Marie Curie* (UPMC), one of the leading science universities in France. In Spring 2012, instructors in two large introductory classes, first-year, second-semester mechanics, and second-year introductory E&M, enrolling approximately 500 and 250 students respectively, introduced ILS into some, but not all of the sections of each class. The specific ILS utilized were Think-Pair-Share questions and Peer Instruction in the main lecture classrooms, and University of Washington Tutorials for Introductory Physics in recitation sections. Pre- and post-instruction assessments (FCI and CSEM respectively) were given, along with a series of demographics questions. Since not all lecture or recitation sections in these classes used ILS, we were able to compare the results of the FCI and CSEM between interactive and non-interactive classes taught simultaneously with the same curriculum. We also analyzed final exam results, as well as the results of student and instructor attitude surveys between classes. In our analysis, we argue that Multiple Linear Regression modeling is superior to other common analysis tools, including normalized gain.

Our results show that ILS are effective at improving student learning by all measures used: research-validated concept inventories and final exam scores, on both conceptual and traditional problem-solving questions. Multiple Linear Regression analysis reveals that interactivity in the classroom is a significant predictor of student learning, showing a similar or stronger relationship with student learning than such ascribed characteristics as parents' education, and achieved characteristics such as GPA and hours studied per week. Analysis of student and instructors attitudes shows that both groups believe that ILS improve student learning in the physics classroom, and increases student engagement and motivation. All of the instructors who used ILS in this study plan to continue its use.

## I. INTRODUCTION

Although the French educational system is different from the American in many ways, the two systems share the goal of helping science students to develop a deep conceptual understanding of their disciplines. A basic grounding in physics is essential to all the sciences, and the principles taught in physics classes (e.g., conservation of energy) touch every science, including physics itself, chemistry, and biology. Students in all these fields take physics, but research has shown that many university science students do not truly understand these basic concepts when they are taught in the traditional lecture style [1,2,3,4,5,6]. Results presented in these references show that using interactive learning strategies (ILS) can significantly improve student understanding of basic science concepts when compared with traditional lecture alone, often by a factor of two or more, and that continued use of these strategies lead to higher gains over time. These interactive learning strategies emphasize creating an environment in which students are active in the classroom, often working collaboratively, and thereby take control of their own learning.

This body of work on the effectiveness of ILS in improving student learning gains in physics classrooms in the U.S., combined with a strong interest in these learning strategies at the *Université de Pierre et Marie Curie* (UPMC) in Paris, France, led us to undertake the study described here. The main focus of this study was to introduce ILS, already shown to successfully improve student learning in the U.S., into the introductory physics classes at UMPC in a systematic way. Though the physics faculty and administrators at UPMC were definitely intrigued by the promise of ILS, they also wanted to see it work in their own educational environment and setting. For example, when this study was first proposed, some UPMC faculty suggested that there were fundamental differences between the cultures of the French and U.S. educational systems that might make the students and instructors resistant to such innovations as ILS. We describe some of these differences in section III. Background and Motivation, below.

We studied two courses: a first-year, second-semester mechanics class, and a second-year E&M course. We we were able to introduce ILS into some sections of each course, but not others, providing two controlled experiments. Student learning gains were assessed using both research-validated concept inventories (FCI and CSEM) and final exam scores. In addition, student demographics information was collected, as was information about both student and instructor attitudes about ILS.

The remainder of this article is divided into the following sections: II. Overview of Interactive Learning Strategies (ILS) used in this Study; III. Background and Motivation; IV. Settings and Participants; V. Study Design; VI. Results; and VII. Conclusions.

## II. OVERVIEW OF INTERACTIVE LEARNING STRATEGIES USED IN THIS STUDY

Implementation of ILS can take many forms, but there are two very common implementations that were used in this study, and which we briefly describe here. The first is *Think-Pair-Share (TPS) questions*, typically used in lecture hall settings (*Amphi* in French), whereby students are asked to answer a multiple choice question designed to test their knowledge of a science concept being presented in the class, first thinking by themselves and choosing an answer (Think), then discussing their answers with their neighbors (Pair), and finally choosing their answer a second time (Share), possibly revising their answer in response to the discussion with their peers (Peer Instruction).  Used together with short lectures on each topic, this ILS is one of the simplest and yet most effective ways for students to actively engage in the classroom [7,8].  The students choices can be collected in various ways: in this study we used a classroom response system (CRS) or "clickers" (*boîtiers réponses* in French)*,* small remote devices that allow an instructor to record the students' answers on their computer, and display them in real time as a histogram.  (The main alternative method of collecting students' answers is flashcards.)

The second major form of interactive learning used in this study is the *tutorial,* primarily used in recitation sections (*Travaux Dirigés* or TD in French). Tutorials, which are done in small groups of 3-4 students, consist of worksheets of questions designed to help students address common difficulties about topics common in introductory physics classes, based on extensive research into the type and nature of these difficulties, and to develop a coherent conceptual understanding on those topics [9,10].  This ILS was pioneered at the University of Washington [11], and has since spread to become one of the leading and most effective ILS used in the United States [10,12].

### III. BACKGROUND AND MOTIVATION

The study took place at *University Pierre et Marie Curie* (UPMC), a major French university located in Paris, France. To provide background to the study and to elucidate the motivation for this study, we begin by briefly describing the French educational system, mostly to contrast it with that of the United States. We then describe the conditions at UPMC that led the physics faculty there to undertake this project of introducing ILS into their classroom, and the associated research study.

The French education system has a number of differences with that in the U.S., and it is beyond the scope of this article to completely describe them. However, we highlight here three major differences that bear on the nature of the participants in our study, and on the cultural differences with students in the U.S. (and to some degree other European countries) in attitudes and motivations of both students and instructors in the system.

In France, students choose an area of study while still in high school (*lycée)*. They can choose between three different streams (*série* in French): natural science, economics and social sciences or literature. Although this choice is not final, changing subjects later is not easy or common. The overwhelming majority of the population studying science in the first year at university comes from the natural science stream of high school, for the simple reason that the scientific background learned in the other streams provides insufficient preparation for the study of science at university.

A second difference between the French and the US system is that in France there is a national system of secondary school education, with a common program of study for each stream. To pursue higher education, students must pass a national exam in their area, the *baccalauréat* (known as the *bac*, for short). The nationally specified programs of study and the existence of a common exam (*bac*) encourages traditional pedagogy that focuses on passing the *bac*, and may have historically limited the flexibility that instructors have in designing their high school courses. It led us to expect that students, and instructors, might resist the introduction of innovative pedagogies, such as ILS.

A third major difference in the French education system concerns the splitting of the student population into tracks in the first years of higher education. After high school, students can follow one of two main tracks, and this choice is primarily based on their grades in high school. Most of the best students attend a post-secondary institution known as *classes préparatoires aux grandes écoles* (*CPGE*), which is a 2-year preparation for competitive entrance exams to enter the *grandes écoles*, the best of which are the highest ranked post-secondary schools in France – these can be thought of as equivalent to the top universities in the U.S. The others students go directly to the *universités*, which are more similar to the US public universities. In the sciences, those two tracks have a roughly equal number of students entering in the first year [13].

UPMC is a top French research university (*université*) with a long-standing international research reputation of the very highest order: in the Shanghai 2012 survey, it is ranked 2nd in France, 8th in Europe and 42nd in the world [14]. However, this research prominence is not reflected in the quality of the undergraduate students. Students who pass the *bac* have the right to attend university, whereas entrance to the *CPGE* and *grandes écoles* is competitive. In addition, universities in France are nearly free to students; in fact, students receive benefits in addition to free tuition, such as discounts for health care, travel, meals, entertainment, etc., which leads some students to enroll at *université* simply to receive these benefits.

In recent years, there has been a growing concern among the physics faculty at UPMC that students were not motivated to learn. The high pass rate for the *bac* (88% in 2011 [15]), together with open enrollment and tracking of weaker students into universities, has led to a low level of success in the first year of university. In France students are given an overall grade for each year of school, and at UPMC, 55% of students fail in their first attempt to pass the first year. Students are often found to be very passive, not only in the lecture hall, but in recitation sections, where they will wait for the instructor to show them answers to the assigned exercises, rather than first attempting to work the problems themselves. Frustration with these problems was one of the main motivating factors for faculty in the Physics Department (*Faculté de Physique*) at UPMC to consider introducing ILS into their classrooms.

The year prior to this study, a few physics faculty members at UPMC had begun to experiment with such strategies, mostly TPS questions in the lecture classroom (*Amphi*). However, there was no systematic, coordinated effort to bring about general change in the practices of the department. After a presentation in France by one of the authors, while visiting from the U.S., on research demonstrating the improved learning gains achieved by students in classes using ILS, a group of instructors, with the support of the department chair (*Directeur de la faculté de physique)*, decided to pursue the study described here: a systematic introduction of ILS into a number of physics classrooms in the first two years at UPMC coupled with a quantitative study of the effectiveness of such ILS in the French university.

## IV. SETTINGS AND PARTICIPANTS

We now describe the system of tracking and majors at UPMC at the time of this study. Unlike the program at the *lycée* (high school), which is standardized nationally, each university designs its own program in each subject. The program in physics (and all the other the sciences) at UPMC has been completely redesigned since this study was conducted, and we describe here the system in place at the time of our study.

In France, the bachelor degree (*licence* in French) is only three years, with increasing specialization as students progress. The years are labeled L1, L2, and L3, where L stands for *licence* and the number indicates level at university. Every student entering UPMC is studying either medicine or science. At the time of this study, the science students in the first year (L1) initially chose to join one of three initial *parcours* (tracks). These tracks are labeled using four letters, where the first two letters indicate the subjects in which students intend to get their *licence* (their major), and the second two letters indicated those other subjects they will study in the track; thus, the main emphasis of the track is indicated by the first two subjects listed. The three tracks are called:

- PCME (*Physique-Chimie-Mécanique-Electronique*), corresponding to Physics, Chemistry, Mechanical Engineering, and Electrical Engineering in the U.S.

- MIME (*Mathématiques-Informatique-Mécanique-Electronique*), corresponding to Math, Computer Science, Mechanical Engineering, and Electrical Engineering, and

- BGPC (*Biologie-Géologie-Physique-Chimie*), corresponding to Biology, Geology, Physics, and Chemistry.

Thus, a student interested in physics or chemistry would join PCME; students interested in math or computer science would choose MIME, and students interested in biology or geology would choose BGPC. Those students who wish to study mechanical or electrical engineering could choose either PCME or MIME, depending on their mathematical ability, or they might choose MIME to avoid studying chemistry, or choose PCME to avoid studying computer science. In the second year (L2), students then choose the particular *licence* (major) they wish to pursue, with the possibility of moving between tracks. Thus, a student in MIME could decide to study physics, since every student studies some physics in the first year (L1).

We now turn to a description of the specific courses we studied. Our study focused on two courses: a first year, second semester mechanics course and a second year electricity and magnetism course [16].  Students in both PCME and MIME study mechanics during the first year in two successive courses, LP111 in the first semester and LP112 in the second semester, but they are divided into different sections of the course based on their track. This "tracking" of enrollment in these courses introduces biases in student abilities between sections that we will return to in the analysis of our results (see section VI. Results).

The division of topics in mechanics between first and second semester at the time of this study was somewhat different from the traditional division in most U.S. colleges and universities. The first semester (LP111) is called "Classical physics I: movement and energy" and focuses on motions of single particles, covering topics such as kinematics and dynamics, energy and work, gravitational and electrostatic forces, and the harmonic oscillator. The second semester (LP112) is called "Classical physics II: dynamics of systems", and focuses, as the name implies, on systems, covering kinematics and dynamics in three dimensions, conservation laws in systems, collisions, statics and dynamics of solids, the two-body problem for central forces, and motion in non-inertial reference frames.

An introduction to electricity and magnetism (E&M) is given in the second year (L2). Many students take this class in the first semester, but there are a number of tracks which do not take E&M until the second semester of the second year. It is these latter students that we studied. There were 4 different E&M classes in the spring semester: LP203-1, LP203-2, LP205 and LE207. These classes serve somewhat different student populations, and are taught in slightly different ways, but the main subject matter is the same, and quite traditional for an introduction to E&M: e.g., conductors, electrostatics (Gauss's Law), magnetostatics (Ampère's Law), and induction (Faraday's Law).

## V. STUDY DESIGN

This study focused on two classes: a first-year, second semester mechanics class (LP112; total enrollment = 476), and a second-year, second semester set of 4 E&M classes (LP203-1,

LP203-2, LP205, and LE207; total enrollment = 264), described above. The study consisted of six main components:

1. *Instructor-training workshops* were held before the semester began, to help faculty learn about best practices in implementing interactive learning in their classroom. The leader of these instructor-training workshops also visited classrooms of instructors to observe and give feedback on implementation when asked, visiting multiple classrooms involved in the study.

2. *Implementation of ILS* in some sections of each class, with varying levels and type of use, creating natural experimental and control groups for each class.

3. *Pre and post-instruction assessment* was done using *concept inventories*, research-validated, multiple-choice instruments designed to measure changes in students' understanding of the basic concepts taught in a course.

4. *Final exam scores* were collected for both the mechanics and E&M classes.

5. *Demographic data* was collected from students via on-line surveys.

6. Both *instructor and student attitudes* were surveyed, on-line for students, on paper or by e-mail for instructors. In addition, the instructors were invited to participate in an end-of-semester debriefing session; the majority attended.

We now describe details of how each of these six study components was implemented.

**1) Instructor-training workshops.** Two training workshops were held in January 2012 for faculty teaching in these two classes, before classes began. These were led by an expert in ILS implementation (Rudolph). The first workshop focused on the implementation of Think-Pair-Share (TPS) questions, including best practices for such implementation, modeled after the workshops developed by the Center for Astronomy Education (CAE) at the University of Arizona [17], but commonly used in classrooms in the United States [7,8]. The second workshop focused on the implementation of tutorials in recitation (*TD*).  This workshop included videos from the *Video Resource for Professional Development of University Physics Educators* [18,19,20], and had faculty work through a sample tutorial. The videos helped participants see common good and bad practices in facilitating student group interactions in completing tutorials. Having instructors experience completing a tutorial themselves allowed them to experience the pedagogical progression of tutorials firsthand, in a setting where they could share and learn from each other's experiences, as well as learn from the workshop leader. Both workshops had about 20 participants.

The workshops were open to all UPMC science and mathematics faculty, and other UPMC faculty besides those teaching in the study courses participated. Most of these instructors participated in one or both of the training workshops; all of the instructors in the two courses in this study who introduced ILS into their classroom participated in both workshops.

**2) Implementation of Interactive Learning Strategies (ILS).**

In the second semester mechanics class, there were five large sections that met in lecture halls (*Amphi*) once a week for two hours, with enrollments ranging from 80-120. The students in these sections then met in recitation sections (*Travaux Dirigés* or *TD*) of 20-30 students, for three 2-hour sessions every two weeks (thus, for an average of three hours per week). Two of the five lecture halls implemented ILS, followed the model of Think-Pair-Share (TPS) questioning, in which shorter, more focused lectures are followed by having the students answer one or more TPS questions [21], while the other three used traditional lectures, mixed in the usual way with examples worked at the board and some demonstrations. In addition, tutorials were used in the recitation sections associated with the classrooms implementing ILS in their lecture halls, but not those in the recitations of the traditional classrooms. Thus, some students in second-semester mechanics were exposed to both TPS questions and tutorials, while others were exposed only to traditional instruction in both the lecture hall and in recitation sections, forming a natural control group for the study.

The two sections of the mechanics class using TPS averaged about eight questions per two-hour lecture class. The instructors in these two sections also used worked examples and demonstrations. However, they occasionally added ILS to their demonstrations by turning them into *predictive demonstrations*. This learning strategy has students use their clickers to make a prediction about the outcome of an experiment or demonstration before it is completed, thereby engaging their thinking in a meaningful way, which greatly improves their comprehension of the physics behind the demonstration [22].

The two lecture sections using ILS also introduced tutorials into their associated recitation sections (*TD*). Traditional instruction in French recitations consists of packets of problems that the students work on throughout the semester. Although the students could work on these problems outside of the recitation classroom, they traditionally do not, and further, most recitation instructors complain that students do not spend time in recitation working on these problems but rather wait for the instructor to present the answers on the board, thinking that possessing these solutions constitute understanding of the material. This passivity of French recitation students was one of the main drivers of the instructors' desire for innovation in their recitations. Hence, the instructors in the five recitations associated with the sections using ILS in their lecture halls each introduced tutorials into some of their recitation sessions. These tutorials were chosen from the University of Washington *Tutorials in Introductory Physics* [11] by the lead instructor in the course, in consultation with the recitation instructors. A total of five tutorials were selected on topics relevant to the material taught in the class, and translated into French. The English titles of these tutorials were: Rotational motion, Newton's second and third laws, Motion in two dimensions, Conservation of momentum in one dimension, and Conservation of angular momentum. Thus, about six hours of the total 34 hours of recitation was spent on tutorials.

The remaining time in the recitations were spent working on the same traditional problems that all the students were assigned in recitation. However, one additional consequence of the introduction of tutorials into some recitations is that, since tutorials are designed to be completed in a group setting, the students in the classes using tutorials began completing the more traditional problems in groups, rather than working individually as was most common in the past.

The second year electricity and magnetism classes also met once a week for two hours in the lecture hall (*Amphi*) and the same for recitations (*TD*). Three of the four sections of the class introduced some level of TPS into their lecture halls, combined with traditional lecture, worked examples, and demonstrations, including predictive demonstrations. However, the level of use of TPS varied between the classrooms considerably [23]. These levels were determined by analyzing feedback received from each instructor at the end of the semester to determine what fraction of their classroom time was spent engaging in ILS, also known as the Interactivity Assessment Score (IAS) [24]. For the most highly interactive class, this score was 0.71 meaning this instructor spent about 70% of his class time on ILS; for the two moderately interactive classes these scores were 0.19 and 0.28, meaning those instructors spent about 20% and 30% of their classroom time on ILS, respectively. One instructor did not use ILS at all, for an IAS of 0. For comparison, a national U.S. study of interactivity in introductory astronomy classes for non-scientists found that 36 instructors had IASs ranging from 0 to 0.47, with a mean of 0.26 [24]. However, the instructors in that study were recruited from participants in training workshops in ILS, and are therefore typical of such participants, not of astronomy or other science instructors generally. A study in the U.S. of the implementation of research-based instructional strategies (RBIS), most of which would be categorized as ILS, found that more that half of all physics instructors they surveyed do not use any RBIS in their classrooms, and therefore have an IAS of zero [25].

    **3) Pre and post-instruction assessment using concept inventories.** To test whether ILS were effective in helping students learn the material in each class, students in both classes were given a concept inventory, a research-validated learning assessment twice: once at the beginning of the semester, before instruction began (pre-instruction), and once at the end, after instruction was complete (post-instruction). By comparing students' scores before and after instruction, it was possible to measure the gain in learning due to the classroom instruction.

    For the mechanics class, the assessment used was the Force Concept Inventory (FCI) [26], a 30-question, multiple-choice instrument developed specifically for use in evaluating student understanding of the basic concepts of Newtonian mechanics which has been shown by rigorous education research to provide a reliable measure of students' learning of basic Newtonian mechanics [1]. We used a French translation of the FCI found on the Arizona State University Modeling Instruction group's legacy research site [27], with minor modifications by two of the French-speaking physicists involved in the study to improve the translation. The content of the FCI is much better matched to the first semester mechanics class (LP111), as is true in many U.S. physics courses, with the notable exception of the topic of Newton's Third Law, which was not covered in-depth until the second semester at UPMC (LP112) at the time of

our study. Nonetheless, we chose to use the FCI in our study, given its ubiquity in the US and elsewhere (allowing us to compare our results with a large number of published results), and the lack of a better instrument. The use of the FCI to assess student conceptual learning in these classes was approved by all the instructors, including those who did not use ILS in their classes.

In the E&M classes, the concept inventory used was the Conceptual Survey of Electricity and Magnetism (CSEM) [28], translated into French by the French-speaking physicists involved in the E&M study. The CSEM contains 32 multiple-choice questions designed to assess students' understanding of the basic concepts of electricity and magnetism. The choice of this concept inventory was made by the four E&M instructors as a group, after a review of existing research-based concept inventories available in the literature, and a discussion of which inventory was the best match of topic and level with the syllabi of the classes.

Both assessments were given on-line, and participation was voluntary. To encourage participation, students in each class were given a small amount of extra credit. Students were informed that their participation would be anonymous, meaning that their results would only be analyzed in aggregate and that whether they participated or not would not affect their grade in the class. The data collection protocol, including informed consent obtained for each participating student, was approved by the IRB at the home institution of the U.S. co-author (Rudolph). In both sets of classes, all the instructors agreed that these concept inventories were reasonable assessments of students understanding of the material taught in the class.

**4) Final exam questions.** In addition to these research-validated assessments, the final exam scores were collected for both the mechanics and the E&M classes. In the mechanics classes, the final exam is common to all sections of the class. One set of questions (about one-third of the exam) based on assessing conceptual understanding was introduced into the exam, while the other two-thirds consisted of more traditional problem-solving questions. In addition to comparing test scores between sections, analysis was done comparing performance on these more traditional test questions to results from the previous year, when every section of the second semester mechanics class used traditional instructional techniques, and when the entire final exam consisted of traditional problem-solving questions.

In the E&M classes, there was one common exercise used on the final exam for all three classes, equal parts conceptual and traditional, allowing comparison between sections of the class. All of the common exam questions were vetted and approved by all of the instructors in each course, who all agreed that they were reasonable measures of student learning, consistent with the learning goals all the instructors (interactive and traditional) shared for the classes.

**5) Demographic data.** For both classes, demographic data were collected on-line in conjunction with the concept inventories. These demographics questions allow us to i) see whether there are any statistical differences in the makeup of the groups being compared (interactive v. traditional sections), and ii) to allow us to probe whether these demographic variables have any effect on student learning, in conjunction with interactivity, via multiple linear regression analysis (see Section VI. Results).

**6) Student and instructor feedback.** For both classes, student and instructor feedback was collected. In the mechanics class, an end-of-semester questionnaire probing students' attitudes towards the course was administered on-line in conjunction with the FCI. This questionnaire asked students to rate their experiences in the class with respect to: 1) their opinion of the instructional style in the class; 2) the learning of both concepts and content of the course; and 3) the effect (if any) of the instructional style of the class on their interest in physics, how hard they worked in class, and the likelihood that they would attend class.

In only one of the second-year E&M classes (LP205) students' attitudes towards the class were collected, with clickers and open response questions on paper, midway through the course. Similar data for all the E&M classes were collected at the end of the semester, but these data were accidentally deleted from the server where they resided, so we only present the mid-term student attitude results here.

Finally, instructors who implemented ILS were given a short questionnaire asking 1) whether they believe the use of ILS improved student learning and assiduity in their class; 2) what motivated them to try ILS in their class; about how they used ILS in their class; and 3) what they liked and disliked about their experience using ILS.

## VI. RESULTS

We now present our results on student learning gains, student feedback, and instructor feedback in the two classes studied: first year, second semester mechanics and second year electricity and magnetism.

### A. Student learning gains

We used multiple measures of student learning gains in the two courses studied. These included research-validated concept inventories (FCI and CSEM for mechanics and E&M respectively [26,28]), common final exam questions given in all sections of each course, and comparisons of exam scores between subsequent years of the mechanics course. All of these measures of student learning consistently show that interactive learning strategies improve student learning compared to more traditional, lecture-only teaching methods.

We begin by presenting evaluation of the concept inventory results, first using the traditional measure of student learning gains for concept inventories: normalized gain. We then go on to discuss the drawbacks of normalized gain, and present what we believe is a superior method of analysis: multiple linear regression modeling. We start the section on multiple linear regression modeling by explaining why it is a superior method for assessing student learning gains; we then describe the demographics surveys that provide additional independent variables for this modeling; and we end by presenting the results of these models for our concept inventory results. We then present multiple linear regression models for common final exam questions in each course, and end the section on student learning gains by presenting a statistical comparison of exam scores between subsequent years of the mechanics class.

### 1. Research-validated concept inventories

(a) *Normalized gain*

The standard measure of student learning gains using concept inventories is normalized gain, $\langle g \rangle = (Post\% - Pre\%)/(100 - Pre\%)$ where Pre% and Post% are the student's percent correct pre-instruction and post-instruction on an assessment instrument [1]. The numerator of this equation is the "Raw Gain" (sometimes referred to as simply "Gain"). The denominator of this equation is designed to remove bias due to unequal starting points for different student populations. Thus, normalized gain is a measure of the fraction of material a student does not already know that he or she has learned in the course. However, there are problems with normalized gain that we detail in the next section. We begin here by presenting our results using traditional normalized gain methods, and then consider alternative methods for assessing student learning on these concept inventories (see next section).

For the second semester mechanics class, we divided the students into two groups: those that used interactive learning in the classroom (PCME21 and 22) and those that did not (MIME21, MIME22, PCME23+). For the interactive sections, the level of interactivity (amount of time spent on TPS questions in lecture hall, and on tutorials in recitation) was roughly the same across all sections.

To assess differences in student learning between the two groups, we began by comparing the average normalized gain for the students in the interactive to those in the non-interactive (traditional) sections of the class using a simple t-test. As can be seen in Table Ia, although the average normalized gain was more than twice as high in the interactive classes, the difference in the mean normalized gain was not statistically significant ($p > 0.05$).

Table I. Tests of statistical significance for normalized gain scores on FCI

| a. Entire FCI | | | | |
|---|---|---|---|---|
| Group | N | %Pre | %Post | $<g>$[†] |
| Interactive | 60 | 53% | 60% | 0.119 |
| Non-interactive | 122 | 49% | 53% | 0.049 |
| *Difference* | | | | *0.071* |
| b. 4 questions of FCI on Newton's Third Law | | | | |
| Group | N | %Pre | %Post | $<g>$[†] |
| Interactive | 39 | 37% | 60% | 0.408 |
| Non-interactive | 72 | 41% | 44% | -0.050 |
| *Difference* | | | | *0.458*[**] |
| *Cohen's D effect size* | | | | *0.583* |

**\*p < 0.05**
**\*\*p < 0.01**
[†] **This column shows the average of the normalized gain for all students in the group, not the normalized gain calculated from the average %Pre and %Post shown in the table. These typically differ due to the non-linearity of normalized gain.**

However, the material assessed by the FCI is primarily taught in the first semester mechanics course at UPMC (as it is in many university physics curricula in the U.S.), so perhaps this result is not surprising. Careful review of the contents of the FCI revealed four questions (Q4, Q15, Q16, Q28) on the concept of Newton's Third Law, which is a central topic of the second semester mechanics class at UPMC. Thus, we calculated Pre%, Post%, Gain, and $<g>$ for these four questions for each student and used a similar t-test to compare the mean $<g>$ for the interactive versus non-interactive (traditional) classes (see Table Ib). Here we find that the students in the interactive sections performed statistically significantly better than those in the traditional, non-interactive sections ($p < 0.01$). The Cohen's-D effect size for this difference is 0.583, indicating a medium-to-large effect size.

Given that we only had four questions to work with in this analysis, we note two effects of this small number of questions:

1) We had to exclude a large number of students (71/182 or 39%) who answered all four questions correctly on the pre-test, since their $<g>$ is undefined (the denominator is zero). Thus, the effect we see is likely enhanced by this exclusion, since we are removing many students whose raw gain is zero or negative. This is a fundamental flaw with $<g>$ that we address in the next section.

2) On the other hand, the small number of questions reduces the sensitivity of the t-test by increasing the effect of the noise in the data, making it more difficult to find statistically significant results. Thus, the fact that we find such a strong statistical difference with such a small N suggests that these results are quite robust.

Though the entire FCI has been validated [26], a subset of only four questions clearly is not. In addition, the relatively low response rate to the FCI (38%), leads to concerns about non-response bias. We acknowledge that these two points limit our ability to interpret these results, in isolation, as strong evidence for the efficacy of ILS in promoting student learning in these mechanics classes. However, these FCI results, when taken as part of the entirety of our results, support the strong evidence we present that ILS did have an overall significant positive impact on student learning in the French physics classrooms we studied.

Table II: Analysis of Variance (ANOVA) of normalized gain scores on CSEM

| Group | N | %Pre | %Post | $<g>^\dagger$ |
|---|---|---|---|---|
| Highly interactive | 42 | 30% | 55% | 0.286 |
| Somewhat interactive | 124 | 29% | 47% | 0.233 |
| Non-interactive | 86 | 28% | 38% | 0.101 |
| *F-Statistic* | | | | *9.28*** |

*$p < 0.05$
**$p < 0.01$
$\dagger$**This column shows the average of the normalized gain for all students in the group, not the normalized gain calculated from the average %Pre and %Post shown in the table. These typically differ due to the non-linearity of normalized gain.**

We now turn to the CSEM normalized gain results for the E&M classes. As noted above in section V. Study Design, the instructors in the E&M classes were surveyed using the Interactivity Assessment Instrument (IAI) of Prather et al. [24] to determine the level of interactivity in each class, the Interactivity Assessment Score or IAS. For the four classes we found IASs of 0, 0.19, 0.28, and 0.71, leading us to define three levels of interactivity: low or non-interactive (IAS = 0), medium or somewhat interactive (IAS ≈ 0.2-0.3), and high or highly interactive (IAS ≈ 0.7). Table II shows the average Pre%, Post%, and normalized gain, $<g>$, for each of these groups: clearly the normalized gain increased as interactivity increased. To test for statistical significance of this result, we compared the normalized gain for these three groups using an analysis of variance (ANOVA) test, and found that the results were highly statistically significant ($p < 0.01$).

(b) *Multiple Linear Regression modeling*

*Why Multiple Linear Regression modeling?* As has been noted by other researchers, there are serious flaws with normalized gain as a measure of learning gain. Wallace & Bailey [29] observe that $<g>$ is not a ratio level variable. A student with twice the normalized gain of another student cannot be said to have learned twice as much since the normalized gain is based on each student's Pre% score. Goertzen et al. [30] noted that normalized gain does not have variance estimates, and often systematically underestimates gains by underrepresented groups, who may start with lower Pre% scores.

In addition to these critiques, normalized gain necessitates the loss of observations where students score a perfect pre-instruction score, because the formula results in the denominator having a value of zero in that case. This does not occur often when a large number of items are used in the testing. However, when a small number of items are used, a perfect pre-score is common. In the evaluation of the four Newton's Third Law questions used here, a full 39% (71 of 182) of the students were eliminated from the analysis for this reason.

Goertzen et al. [30] account for some of these issues by analyzing Pre%, Post%, and raw gain for the FCI at the group level. This successfully accounts for different starting points for individual subgroups within the population. However, this approach is also not without its flaws. First, it presents the gains or losses in learning at the group level, which masks the learning gains and losses at the individual level. Second, this approach, by dividing the sample into subgroups, limits the number of observations included for each subgroup, and thus reduces the statistical power (sample size N) of the analysis. Third, this method of analysis only accounts for one independent variable at a time. To allow analysis of the effect of multiple variables, one could create subgroups based on many such factors, but that would only further reduce the statistical power of the analysis for each variable (by reducing N), and would thus require a very large sample. Finally, Goertzen et al.'s [30] approach only works on two-level variables, so analysis of a continuous variable (such as GPA) would have to be reduced to two groupings (e.g., low and high), thereby throwing away information, subjecting the analysis to the researcher's particular choice of categories, and reducing the analytic potential of the results.

Multiple linear regression modeling is a statistical method that allows many independent variables to be fitted simultaneously to measure the relative effect of those variables on a single dependent variable. Thus, each independent variable's effect is isolated from the others, thereby controlling for those other variables.

The use of multiple linear regression modeling addresses many of the issues with normalized gain and the analysis of Goertzen et al. [30] identified above:

- Regression analysis is conducted at the individual level, thus focusing on the effect of various factors on individual learning

- It allows the researcher to incorporate many independent variables into the analysis at one time with a minimal reduction of statistical power

- It controls for these independent variables thus isolating the effect of interactive learning separate from other factors that might influence an individual's learning in the class

- It permits variables of all levels of measurement (nominal, rank, interval, and ratio) to be incorporated into the models, rather than reducing the level of measurement to only two groups (dichotomous) as done by Goertzen et al. [30]

- The inclusion of each individual's pre-instruction score into the model as an independent variable allows one to control for the effects the pre-instruction score has on post-instruction score

- Regression analysis can be performed with sample sizes considerably smaller than the subgroup analysis method demands

- In addition, the relative effect sizes of all independent variables can be measured against each other thus allowing us to determine the absolute and relative strengths of each independent variable

We now describe the demographics surveys we conducted to allow us to use such demographic variables in our regression analysis, and then present the regression analysis itself for each of the classes in our study.

*Demographics surveys.* To help understand the nature of the student population in our study, and to aid in probing the effect of demographics (along with ILS) on student learning gains using linear regression, we administered an on-line demographic survey to each class. For the mechanics class, this consisted of a series of 15 questions including both ascribed characteristics (e.g., gender, French as a native tongue, level of education of each parent), and achieved characteristics (e.g., year and type of *baccalaureat* (end-of-high-school exam), GPA in the first semester of university, hours per week spent studying). To look for demographic differences between the interactive v. non-interactive (traditional) sections, we coded each question and ran t-tests for differences between the populations. We found that the two groups were statistically indistinguishable with the exception of characteristics related to the tracking inherent in the French system, namely year and type of *baccalaureat*. No statistically significant differences were found for the ascribed characteristics, or in first semester GPA, or hours spent each week studying for the class.

For the E&M classes, the demographics survey consisted of 20 questions, very similar to those used in the mechanics class. The main differences were that students were asked for the GPA in both semesters of their first year (L1), allowing us to construct an overall first year GPA, and students were asked how many physics courses they had taken in their first year. To compare if the different classes had differing demographics, we regrouped the E&M students into two groups, those with any interactivity in their class (medium or high interactivity) and those with no interactivity in their class (low interactivity). Comparing these two groups' demographics using t-tests for each demographic variable showed no statistically significant differences, other than the year they completed their *baccalaureat*, and the number of physics classes they had taken in their first year of university (L1), both of which are due to the tracking of students at UPMC. Again, no statistically significant differences were found for any ascribed characteristics, or in first year GPA, or hours spent each week studying for the class.

*Multiple linear regression modeling.* For both classes, we constructed a series of linear regression models in which we successively added independent variables, to isolate the effect of adding different variables to each model. Table III shows the results of a series of three models using the data for the mechanics class, with FCI Newton's Third Law Gain (based on the four FCI questions described above) as the dependent variable [31]. The first column for each model lists the coefficient of each independent variable, with one or two asterisks indicating if that variable statistically significantly predicts the dependent variable at the $p < 0.05$ (*) or $p < 0.01$ (**) level. The second column for each model shows the standardized coefficient for each independent variable, which is the coefficient in units of standard error. These latter measures, unlike the coefficients, are scale independent, and therefore allow the direct comparison of size of the relationship between independent variables (the standardized coefficient is equivalent to Cohen's-D effect size in a single variable t-test). At the bottom of each model is indicated the F-value of the model labeled with asterisks to indicate the level of statistical significance of the entire model, plus the sample size, N, and the adjusted R-squared of the model; this last value is a measure of what fraction of the variance in the dependent variable is accounted for by the model.

**Table III. FCI Newton's Third Law -- Models 1-3**

| Independent Variable | Dependent variable = FCI Newton's Third Law Gain | | | | | |
|---|---|---|---|---|---|---|
| | 1 | | 2 | | 3 | |
| | Coefficients (standard error) | Standardized Coefficients | Coefficients (standard error) | Standardized Coefficients | Coefficients (standard error) | Standardized Coefficients |
| Constant | 1.262** (0.313) | | 1.022* (0.460) | | 0.721 (0.460) | |
| Male | 0.134 (0.300) | 0.043 | -0.055 (0.299) | -0.017 | -0.141 (0.292) | -0.045 |
| Parents' Education | -0.043 (0.047) | -0.084 | -0.054 (0.046) | -0.106 | -0.034 (0.045) | -0.067 |
| FCI Newton's Third Law Pre-score | -0.452** (0.096) | -0.446** | -0.502** (0.094) | -0.495** | -0.480** (0.092) | -0.474** |
| First semester Mechanics final exam | | | 0.033** (0.012) | 0.258** | 0.035** (0.012) | 0.274** |
| Hours studied per week | | | -0.041 (-0.041) | -0.158 | -0.042 (0.023) | -0.160 |
| Level of course interactivity | | | | | 0.701** (0.266) | 0.231** |
| F Value | 7.75** | | 6.82** | | 7.19** | |
| N | 102 | | 102 | | 102 | |
| Adjusted R-Square | 0.167 | | 0.224 | | 0.269 | |

*p < 0.05
**p < 0.01

The first model includes only ascribed characteristics (gender and parents' education), plus FCI Pre-score as independent variables. This model is statistically significant, only due to the expected negative correlation between Pre-score and Gain, with an adjusted R-squared of 0.167.

The second model adds achieved characteristics, namely the students' scores on the first semester mechanics final exam and number of hours spent studying each week; only the first of these two statistically significantly predicts the FCI Newton's Third Law Gain.

This second model has an adjusted R-squared of 0.224, a 34% increase over model one, indicating that, perhaps not surprisingly, scoring well on the first semester final exam strongly predicts learning Newton's Third Law. It might seem surprising that we did not find any relationship between the number of hours studied per week and *conceptual* learning of Newton's Third Law.

The third model introduces level of course interactivity as an independent variable, which is found to be highly statistically significant at the $p < 0.01$ level. The adjusted R-squared of this final model continues to increase to 0.269, indicating that the single variable of interactivity contributes significantly to the predictive power of the model. It is striking that the standardized coefficient for interactivity is comparable in size to that of the first semester final exam score, suggesting that level of interactivity in the class has a similarly large effect as how well a student performed on a final exam designed to test their knowledge of first semester mechanics (see Figure 1).

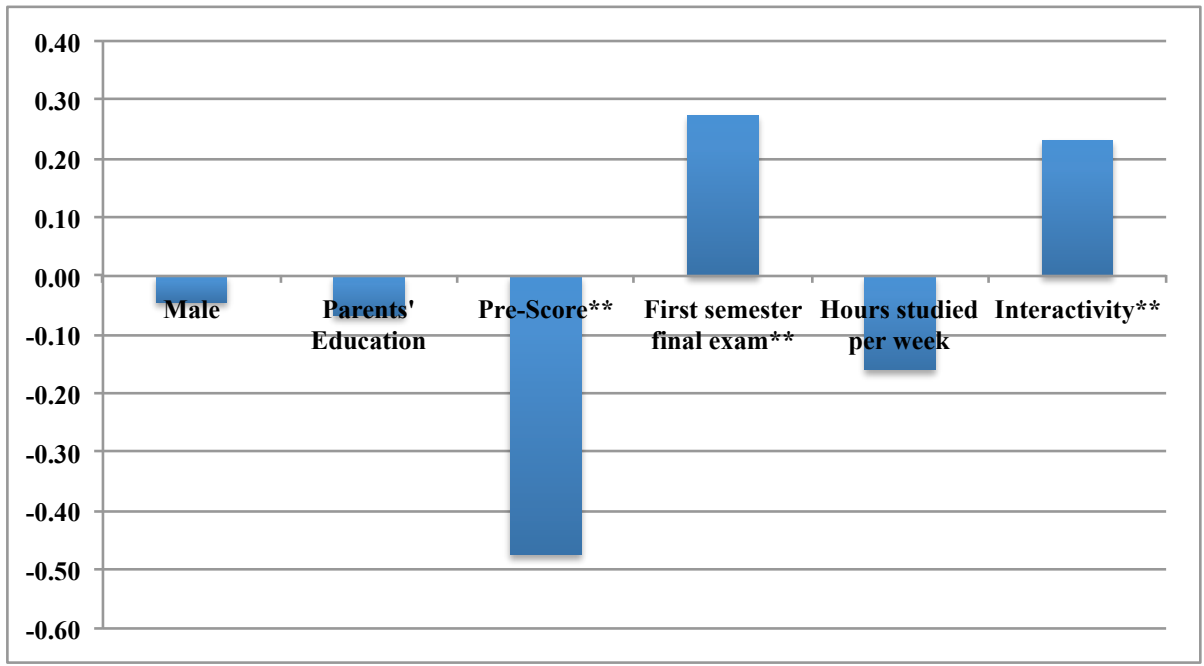

**Figure 1 Standardized coefficients for Model 3 with FCI Newton's Third Law Gain as the dependent variable (from Table III).**
 **\*p < 0.05**
**\*\*p < 0.01**

Table IV shows the results of a similar series of three models using the data for the four E&M classes, with CSEM Gain as the dependent variable.  Again, the first model includes only ascribed characteristics (gender, parents' education), and again, only the Pre-score is statistically significant of these three independent variables, with an adjusted R-squared of only 0.114.  In this series, the second model adds the achieved characteristics of the students' overall first year (L1) GPA, an average of their first two semesters' GPA, and number of hours studied per week. In the case of the CSEM, neither of these variables is statistically significant, and the R-squared is essentially unchanged (0.120). The third model adds interactivity level, coded as 0, 1, and 2, for low (none), medium, and high interactivity, respectively. The level of interactivity is highly statistically significant ($p < 0.01$), and the R-squared jumps *73%* from the addition of this single variable, to 0.208, suggesting that interactive learning was strongly related to student learning of the material in the CSEM.

**Table IV. CSEM -- Models 1-3**

| Independent Variable | 1 Coefficients (standard error) | 1 Standardized Coefficients | 2 Coefficients (standard error) | 2 Standardized Coefficients | 3 Coefficients (standard error) | 3 Standardized Coefficients |
|---|---|---|---|---|---|---|
| | Dependent variable = CSEM Gain | | | | | |
| Constant | 9.802** (1.404) | | 5.995 (3.692) | | 5.813 (3.501) | |
| Male | -0.723 (0.956) | -0.058 | -0.407 (0.990) | -0.033 | 0.160 (0.948) | 0.013 |
| Parents Education | 0.243 (0.155) | 0.123 | 0.226 (0.156) | 0.114 | 0.175 (0.149) | 0.089 |
| CSEM Pre-Score | -0.645** (0.143) | -0.356** | -0.668** (0.147) | -0.369** | -0.717** (0.139) | -0.396** |
| First year overall grade | | | 0.194 (0.259) | 0.060 | 0.146 (0.246) | 0.045 |
| Hours studied per week | | | 0.114 (0.081) | 0.112 | 0.070 (0.077) | 0.069 |
| Level of course interactivity | | | | | 2.535** (0.604) | 0.313** |
| *F Value* | *7.585** | | *5.166** | | *7.716** | |
| *N* | *154* | | *154* | | *154* | |
| *Adjusted R-Square* | *0.114* | | *0.120* | | *0.208* | |

*p < 0.05
**p < 0.01

As seen in Figure 2, the level of interactivity is the dominant factor in predicting a students' gain on the CSEM, other than their pre-score, with a standardized coefficient of about 0.3 (between a small and medium effect).

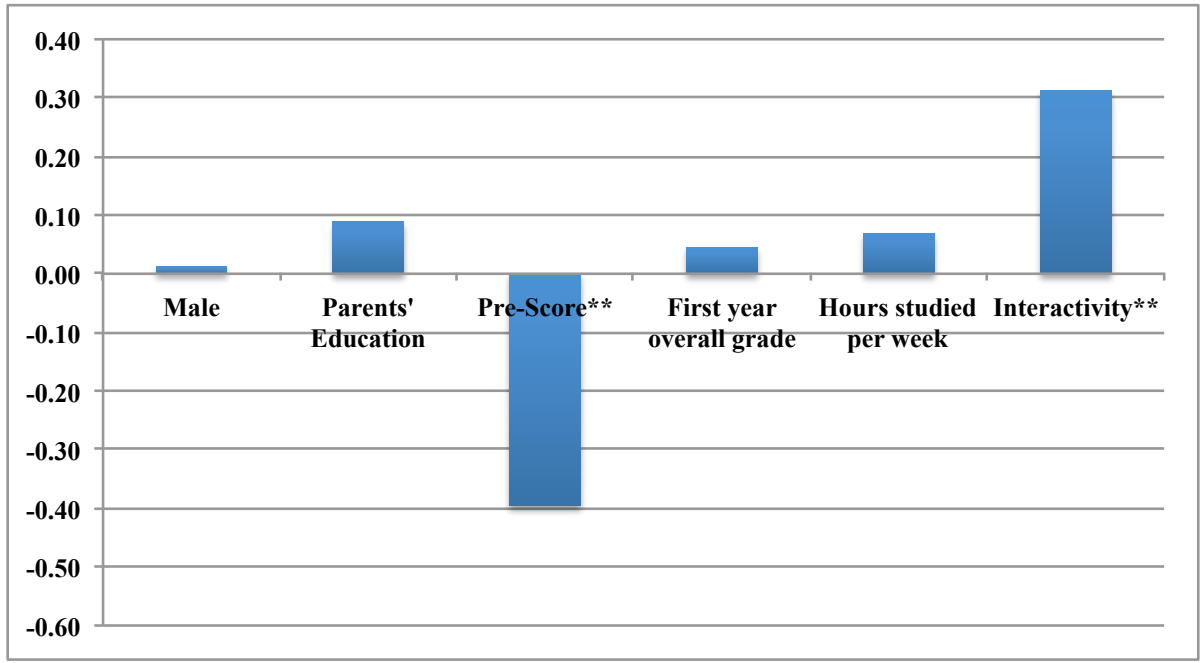

**Figure 2. Standardized coefficients for Model 3 with CSEM Gain as the dependent variable (from Table IV).**
 **\*p < 0.05**
**\*\*p < 0.01**

In summary, interactivity was a dominant factor in models for both concept inventories used to assess student learning in these two physics classes: the 4 questions in the FCI on Newton's Third Law in the second semester mechanics class, and the entire 32-question CSEM for the second year E&M classes. These results reaffirm similar results seen in large-scale studies of the effect of interactivity in US classrooms [1,32], confirming that improvements in conceptual learning of physics concepts can take place in the French university system.

*2. Common final exam questions*

To further probe the effect of interactive learning on student learning, common final exam problems were administered in both classes. In the second semester mechanics class, the sections are all part of a centrally administered class, and the entire final exam is always common, and typically divided into three roughly equally weighted parts. In past years this exam has been entirely made up of traditional problem-solving questions. However, in the semester studied here, one set of questions, about one-third of the exam, was designed to probe conceptual understanding, and the other two sets of questions were the more traditional, problem-solving questions. In the E&M classes, the classes are traditionally taught independently, so final exams are usually not common. However, as part of this study, the instructors of these classes voluntarily agreed to include one set of common problems, equaling about one-third of the exam: these common problems were roughly half conceptual in nature and half traditional, problem solving questions. We now present an analysis of these exam results using the linear regression techniques outlined in the previous section.

For the mechanics class, we constructed a series of linear regression models with score on the common conceptual final exam questions as the dependent variable, shown in Table V. The first model used only ascribed characteristics (gender, parents' education) and found that gender was weakly statistically significant ($p < 0.05$) and that the overall model was statistically significant, but with an adjusted R-squared of only 0.016. In the second model we added achieved characteristics: first semester mechanics final exam score and hours studied per week. The statistical significance of gender disappeared, and both of the achieved characteristics were statistically significant: $p < 0.01$ for the first semester final exam score and $p < 0.05$ for hours studied per week. Together, addition of these two variables significantly improved the predictive power of the model, raising the adjusted R-squared to 0.212. It is perhaps not surprising that these two achieved characteristics would correlate with performance on the conceptual final exam problems, particularly performance on the final exam from the first semester mechanics course. It is worth noting that hours studied per week was significant (though weakly) in predicting performance on a set of conceptual final exam problems, but not in predicting performance on the Newton's Third Law problems of the FCI.

**Table V. Mechanics Common Final Exam -- Models 1-3**

| Independent Variable | Dependent variable = Mechanics Common Conceptual Final Exam Problems | | | | | |
|---|---|---|---|---|---|---|
| | 1 | | 2 | | 3 | |
| | Coefficients (standard error) | Standardized Coefficients | Coefficients (standard error) | Standardized Coefficients | Coefficients (standard error) | Standardized Coefficients |
| Constant | 2.946** (0.374) | | 0.081 (0.572) | | -0.232 (0.572) | |
| Male | 0.849* (0.410) | 0.117* | 0.602 (0.376) | 0.117 | 0.514 (0.370) | 0.100 |
| Parents' Education | 0.052 (0.066) | 0.063 | 0.008 (0.059) | 0.010 | 0.027 (0.058) | 0.033 |
| First semester Mechanics final exam | | | 0.087** (0.016) | 0.401** | 0.087** (0.016) | 0.403** |
| Hours studied per week | | | 0.068* (0.032) | 0.153* | 0.069* (0.032) | 0.158* |
| Level of course interactivity | | | | | 0.966** (0.353) | 0.193** |
| *F Value* | *2.272** * | | *11.535** * | | *11.114** * | |
| *N* | *158* | | *158* | | *158* | |
| *Adjusted R-Square* | *0.016* | | *0.212* | | *0.244* | |

**\*p < 0.05**
**\*\*p < 0.01**

The third model added interactivity level in the class, and again the adjusted R-squared of the model increased (modestly) to 0.244, and interactivity level was found to be highly statistically significant ($p < 0.01$) at predicting performance on the common conceptual final exam questions. The standardized coefficient for interactivity, though not as high as that for performance on the first-semester final exam, was similar to that of hours studied per week, suggesting that introducing interactivity into a classroom can have a comparable impact on student learning to the number of hours a student studies per week (see Figure 3).

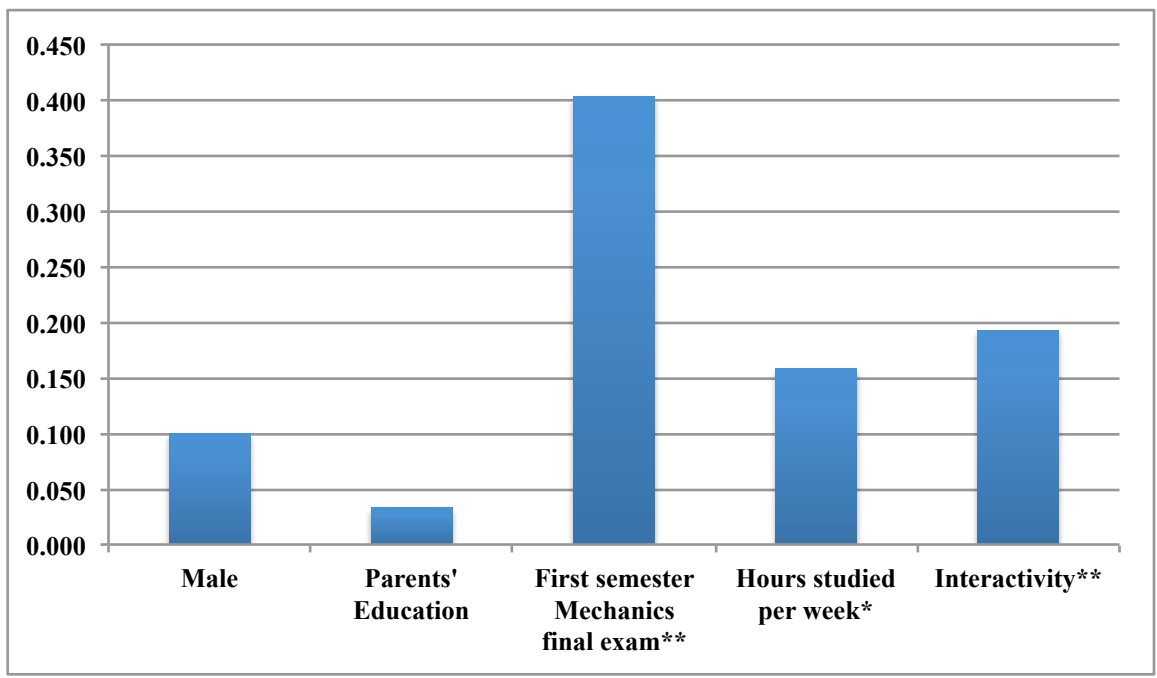

**Figure 3. Standardized coefficients for Model 3 with Mechanics Common Final Exam Problem score as the dependent variable (from Table V).**
*p < 0.05
**p < 0.01

Analysis of the effect of interactivity on student performance on the common traditional problems of the mechanics final exam is more complex, due to biases in student ability introduced by tracking into the course, and is postponed to the next section VI.A.3.

**Table VI. E&M Common Final Exam -- Models 1-3**

| Independent Variable | Dependent variable = E&M Common Final Exam Problems | | | | | |
| --- | --- | --- | --- | --- | --- | --- |
| | 1 | | 2 | | 3 | |
| | Coefficients (standard error) | Standardized Coefficients | Coefficients (standard error) | Standardized Coefficients | Coefficients (standard error) | Standardized Coefficients |
| Constant | 29.350** (3.794) | | -30.428 (20.780) | | -29.354 (19.513) | |
| Male | -1.112 (5.567) | -0.017 | 1.266 (5.531) | 0.019 | 3.890 (5.517) | 0.059 |
| Parents' Education | 3.051** (0.901) | 0.288** | 2.777** (0.875) | 0.262** | 2.347** (0.842) | 0.221** |
| First year overall grade | | | 3.265* (1.394) | 0.194* | 2.880* (1.332) | 0.171* |
| Hours studied per week | | | 1.169** (0.445) | 0.221** | 0.887* (0.430) | 0.167* |
| Level of course interactivity | | | | | 11.820** (3.178) | 0.301** |
| *F Value* | *5.862** | | *6.703** | | *8.675** | |
| *N* | *131* | | *131* | | *131* | |
| *Adjusted R-Square* | *0.070* | | *0.149* | | *0.228* | |

**\*p < 0.05**
**\*\*p < 0.01**

For the common final exam problems used in the four E&M courses, we constructed another set of three linear regression models, with the common final exam problem scores as the dependent variable (see Table VI). Recall that these final common exam questions consisted of half conceptual questions, and half traditional problem-solving questions. The first model, with ascribed characteristics of gender and parents' education, found the latter to be highly statistically significant ($p < 0.01$), the only model to find such a relationship, but with an overall adjusted R-squared of only 0.070. The second model added the achieved characteristics of first year overall grade and hours studied per week, and found both of these variables to be statistically significant: first year overall grade at a lower level ($p < 0.05$) than hours studied per week ($p < 0.01$). Parents' education continued to be statistically significant in this second model ($p < 0.01$). This second model more than doubled the adjusted R-squared to a still modest 0.149.

In the final (third) model, level of course interactivity was added and was again found to be highly statistically significant ($p < 0.01$) and the adjusted R-squared jumped an additional *50%* to 0.228, strong evidence that interactivity had a large impact on student learning. All of the previously statistically significant variables remained significant, though hours studied per week had a lower significance in Model 2 ($p < 0.05$) than in previous models.

In addition, a comparison of the standardized coefficients of the variables in Model 3 shows that interactivity was *the single most important variable* in predicting student success on the common final exam questions, both conceptual and traditional (see Figure 4). The effect of interactivity was larger than parents' education, first year grade (GPA), and hours studied per week, all measures that would traditionally be considered strong predictors of students success, but none of which is under the instructor's control. Thus, we consider these results to be the strongest we found for a beneficial effect of interactivity in promoting student learning.

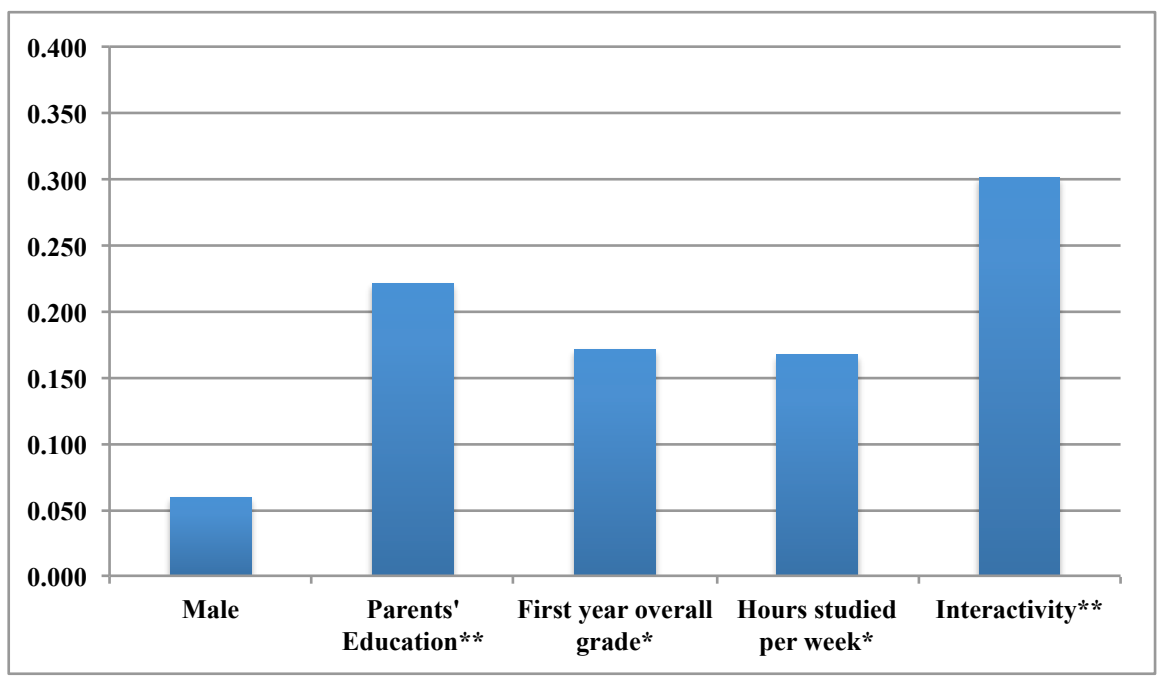

**Figure 4. Standardized coefficients for Model 3 with E&M Common Final Exam Problem score as the dependent variable (from Table VI).**
 *p < 0.05
**p < 0.01

### 3. Comparison of exam scores between years

We demonstrated in the previous section that in the first-year, second semester mechanics class, class interactivity level was a statistically significant predictor of student performance on a set of *conceptual* common final exam questions. Though this result does provide corroboration of the result showing a similar statistically significant relationship of class interactivity with Newton's Third Law questions on the FCI, both of these measures of student learning are conceptual in nature, and one might reasonably ask if interactivity influences performance on more traditional exam problems as well. We should note that we did see a very strong correlation of student learning with interactivity in the E&M class on common final exam questions including *both* conceptual and traditional exam problems, but nonetheless, we wished to investigate the effect of interactivity on performance on traditional exam problems independently in the mechanics class. Unfortunately, this investigation is complicated by the student tracking found in the French university classrooms, in particular in the non-random assignment of students to classes at UPMC: the students in two of the three non-interactive sections come from the MIME track, which consists of students who are traditionally stronger than those in the interactive sections (PCME 22/23), as evidenced by final exam scores in previous years. Table VIIa lists the Spring 2011 scores on the final exam scores, which consisted entirely of traditional problem-solving questions, by tracking group (*parcours*); it is clear that the MIME students perform at a higher level than the PCME 21/22 students, which in turn perform higher than the PCME 23+ students, a pattern seen over several years.

Table VIIb shows a similar breakdown of final exam scores for Spring 2012, the term studied here, in which the PCME 21/22 classes were taught interactively. The only change in the course was the introduction of interactive learning into those two sections of the class. Clearly, the MIME sections still outperform the PCME 21/22 sections on the total exam score, but by a smaller amount [33].

**Table VIIa. Spring 2011 Mechanics Final Exam Scores**

|  | N | Score[†] | Score/MIME |
|---|---|---|---|
| **MIME** | 202 | 28.1 | 1.00 |
| **PCME 21/22** | 157 | 17.3 | 0.61 |
| **PCME 23+** | 112 | 16.6 | 0.59 |
| **ALL** | 471 | 21.7 | |

**Table VIIb. Spring 2012 Mechanics Final Exam Scores**

|  | N | Score[†] | Score/MIME | Conceptual Qs | Traditional Qs |
|---|---|---|---|---|---|
| **MIME** | 185 | 17.0 | 1.00 | 3.4 | 13.6 |
| **PCME 21/22*** | 149 | 14.4 | 0.77 | 4.0 | 10.5 |
| **PCME 23+** | 104 | 11.4 | 0.66 | 2.5 | 8.8 |
| **ALL** | 438 | 14.6 | | 3.4 | 11.2 |

[†]Scores shown here are out of a possible 55 points (corresponding to the percent of the total grade determined by the final exam). These absolute scores vary from year-to-year due to differences in grading and cannot be simply compared between years without some normalization.
*Interactive sections

The last two columns of Table VIIb show the breakdown of exam scores into the conceptual questions analyzed in the previous section, and the remaining two sections of the exam, which consisted of traditional problem-solving questions such as were found on the Spring 2011 exam. Here one notes that the interactive sections outperformed the traditionally taught MIME sections on the conceptual questions, in spite of the traditionally better performance of those latter sections on the overall exam and course. This is consistent with our findings from the previous section.

We note that, though the interactively taught PCME 21/22 sections in Spring 2012 do not perform as well as the traditionally taught MIME sections on the traditional problem-solving questions of the exam, the gap has closed somewhat, and one might wonder if the interactive students had performed better on these traditional exam problems, *relative to their usual performance.* To determine if this last hypothesis were true, we performed the following analysis:

1. For each term (Spring 2011 and 2012), we calculated the mean and standard deviation of the MIME sections' exam scores on the traditional problem-solving questions on each exam (the entire exam in Spring 2011, two-thirds of the exam in Spring 2012).

2. We used this mean and standard deviation (SD) for each term to construct normalized Z-scores for each PCME 21/22 student in both terms, applying the normalization within the appropriate term using the formula:

   Z-score = [(Raw exam score)-(Mean MIME score)]/(SD of MIME)

3. The distribution of Z-scores for Spring 2011 and 2012 were then compared via a t-test of significance of the difference in means in the usual way.

This analysis makes the assumption that the relative level of the students in the various tracks (MIME and PCME 21/22) remains the same from year-to-year, and that the only change in the PCME 21/22 classes is the introduction of interactivity in Spring 2012.

**Table VIII. Z-score Comparison of PCME 21/22 Mechanics Final Exam Scores[†]**

| YEAR | N | Mean | Std. Dev. |
|------|------|------|-----------|
| **2011** | 137 | -0.70 | 1.12 |
| **2012** | 148 | -0.41 | 0.88 |
| **Difference** | | 0.29** | |

*p<0.05
**p<0.01
[†]Traditional problem-solving questions only

Table VIII shows the mean and standard deviation of the Z-scores of the PCME 21/22 sections, and the difference. The t-test showed a highly statistically significant difference ($p < 0.01$) between these means, suggesting that adding interactivity in Spring 2012 did improve the performance of students on traditional problem-solving questions on the final exam, relative to the performance of their peers in the previous year. Because the mean is already presented in units of standard deviation, the mean difference, 0.29, is equal to the Cohen's-D effect size, suggesting a small-to-medium effect.

To summarize the results of the student learning gain results for this study: we find that on all measures of student learning gain, research-validated concept inventories and common final exam questions, both conceptual and traditional problem-solving, the introduction of interactive learning into the French university physics classroom had a statistically significant positive impact on student learning. The effect sizes vary, but are often quite large, and are often larger than other student characteristics that we would expect to have a large impact on student learning, such as GPA or hours studied per week.

## B. Student attitude results

To study French students' attitudes towards interactive learning, students in both classes were given attitude surveys. In the mechanics class, students were given an on-line voluntary attitude survey at the end of the semester, in conjunction with the FCI post-test. The results of this survey show a clear difference in assessment of the course by students enrolled in sections using interactive and traditional instruction.

Table IX shows the six questions asked of students in both types of classrooms (interactive and traditional) concerning their opinion: 1) of the instruction in their classroom; 2 and 3) of the impact of that instruction on their learning of content of the course and of the concepts; 4 and 5) of the impact of the instructional style on their interest and assiduousness in class; and 6) of the impact on their attendance. The answers to these questions were converted to a 5-point Likert scale, where 5 represented a large positive impact, and 1 represented a large negative impact; 3 was neutral. The mean of the responses to each question was calculated, and percentages were calculated for positive responses (responses 4 and 5, labeled +), neutral responses (response 3, labeled 0), and negative responses (responses 1 and 2, labeled −). As can be seen in the table, students in the interactive sections of second semester mechanics rated the course higher on all six elements of the questionnaire, with increases in mean scores ranging from 12% to 26%. A t-test of the difference in means for these six questions found that all of these differences were statistically significant (the p-values are listed for each question in the table). The last column shows the Cohen's-D effect size for each question: these vary from 0.26 to 0.53, in the small-to-medium range of effect size.

**Table IX. Student Year-end Attitudes in Second Semester Mechanics**

| | Question | Traditional | | | | Interactive | | | | Statistics | | |
|---|---|---|---|---|---|---|---|---|---|---|---|---|
| | | N | average score | | | N | average score | | | % increase | p value | Cohen's-D effect size |
| 1 | Your general opinion of the teaching in LP112 is | 132 | 3.06 | + | 41.7% | 74 | 3.42 | + | 52.7% | 11.8% | 0.0144 | 0.36 |
| | | | | 0 | 27.3% | | | 0 | 32.4% | | | |
| | | | | − | 31.1% | | | − | 14.9% | | | |
| 2 | To what extent would you say that the way in which teaching took place in LP112 promoted or otherwise impeded the learning of contents? | 131 | 2.93 | + | 39.7% | 73 | 3.69 | + | 63.0% | 25.9% | 0.0003 | 0.53 |
| | | | | 0 | 23.7% | | | 0 | 15.1% | | | |
| | | | | − | 36.6% | | | − | 21.9% | | | |
| 3 | To what extent would you say that the way in which teaching took place in LP112 promoted or otherwise impeded the understanding of the concepts? | 131 | 3.02 | + | 43.5% | 74 | 3.59 | + | 62.2% | 18.9% | 0.0004 | 0.52 |
| | | | | 0 | 21.4% | | | 0 | 23.0% | | | |
| | | | | − | 35.1% | | | − | 14.9% | | | |
| 4 | To what extent would you say that the way in which teaching took place in LP112 has increased or decreased your interest in physics? | 132 | 2.96 | + | 33.3% | 74 | 3.32 | + | 43.2% | 12.2% | 0.0137 | 0.36 |
| | | | | 0 | 37.1% | | | 0 | 35.1% | | | |
| | | | | − | 29.5% | | | − | 21.6% | | | |
| 5 | To what extent would you say that the way in which teaching took place in LP112 encouraged you to work hard in your course? | 132 | 2.69 | + | 23.5% | 74 | 3.04 | + | 35.1% | 13.0% | 0.0142 | 0.36 |
| | | | | 0 | 39.4% | | | 0 | 44.6% | | | |
| | | | | − | 37.1% | | | − | 20.3% | | | |
| 6 | To what extent would you say that the way in which teaching took place in LP112 encouraged you to attend class? | 131 | 2.75 | + | 24.4% | 73 | 3.27 | + | 41.1% | 18.9% | 0.0022 | 0.45 |
| | | | | 0 | 35.1% | | | 0 | 42.5% | | | |
| | | | | − | 40.5% | | | − | 16.4% | | | |

We highlight two other conclusions from this table. First, it is particularly striking that the two greatest differences, 26% and 19%, came on questions about students' perceptions of the improvement in their *learning*, either factual knowledge or concepts, i.e., students believe that they learn better with ILS. Second, the number of students having a negative opinion of the course (those who chose 1 or 2 for question 1) is half as large with ILS, decreasing from a 30% for traditional sections to 15% in the interactive sections.

**Table X. Midterm Student Attitudes in One Second Year E&M Class (LP205)**

| | Question | Interactive | | | |
|---|---|---|---|---|---|
| | | N | average score | | |
| 1 | What is your general opinion on interactive learning? | 35 | 4.21 | + | 91% |
| | | | | 0 | 6% |
| | | | | − | 3% |
| 2 | To what extent would you say that interactive learning promoted or otherwise impeded the understanding of the concepts? | 27 | 4.05 | + | 92% |
| | | | | 0 | * |
| | | | | − | 8% |
| 3 | To what extent would you say that interactive learning has increased or decreased your interest for physics? | 36 | 4.06 | + | 61% |
| | | | | 0 | 31% |
| | | | | − | 8% |
| 4 | To what extent would you say that the way in which teaching took place encouraged you to work hard in your course? | 34 | 4.35 | + | 82% |
| | | | | 0 | 15% |
| | | | | − | 3% |

*Because this was a mid-term feedback, this question included an answer "I don't know yet, I am waiting for the results of the exams", which 9 students chose. These answers were not included in the calculation of the average score.

A midterm student attitude survey was also administered for the E&M students in the LP205 class only. The results of this survey, summarized in Table X, show a very positive reaction of students to interactive learning. Of particular note is the responses to Question 2, "To what extent would you say that interactive learning promoted or otherwise impeded the understanding of the concepts?" Because this was a midterm feedback, the responses to Question 2 included an option, "I don't know yet, I am waiting for the results of the exams", which was selected by 9 (25%) of the students. Of the students who did not choose this option, over 90% indicated that interactive learning promoted their understanding of the concepts of the class. Even including all the students in the results, 69% of the students selected a positive response that indicates that interactive learning promoted their understanding of the concepts of the class.

While the instructor in this class observed no significant increase in class attendance compared to previous years, a very high proportion (82%) of the students who did attend class stated that interactive learning increased their assiduousness in class.

### C. Faculty attitude results

At the end of the semester, the instructors that introduced interactive learning into their classroom participated in an end-of-semester meeting to debrief their experiences. In addition, they were invited to complete a survey on their experiences; 100% (N=15) of the instructors complied. The first two questions asked them to give an overall score, on a Likert scale (5=a great deal, 1=not at all), for 1) the effectiveness of interactive learning in the improvement of student learning and 2) student motivation to be more active and diligent in the class. The average score for these two questions was 3.8 (N=15) and 3.6 (N=14) respectively, indicating that overall the instructors felt that ILS had improved learning and student motivation in their classes. Ten instructors (about two-thirds) chose 4 or 5 for each question.

Instructors were also asked what they particularly liked about the use of interactive learning. Many of these responses highlighted the well-known impact of ILS in increasing student participation in class, and of providing feedback to both students and instructors about student understanding. In regards to the former, instructors commented that, "sessions were more interactive", "I like the interaction with students; I like that they are encouraged to participate", and "I find ILS help to establish a much better communication between the teacher and students and also between students themselves". Instructors who commented on the improved feedback said, "I had a real sense of what is really going on for the students, whether they are understanding or not", and "For me, I understood the gap between where the students were in their learning and what we had to do".

In addition to these usual benefits of ILS, the answers also raised a few points that address the traditional issues of the French educational background. For example, one instructor noted that ILS create a "possibility of a different way to present the concepts, through questions and examples instead of demonstrations". It is therefore accompanied with "less mathematical background" and some instructors were "satisfied to reveal with the questions the link between physical concepts and their use in everyday life situations". Another point raised by the instructors is the "near miracle" of having a "student explain his reasoning in front of their fellows during lectures", since typically, French university students are quite passive and reluctant to answer questions from the instructors in a traditional lecture. One instructor also pointed that it is "much more fun to hear a student give the correct argument than doing it yourself". Finally, we note that one instructor (who is not French), who had expressed extreme skepticism about whether French students (and instructors) would accept ILS into the classroom, made the following comment: "At the outset, for various cultural and other reasons, I mentioned that the method would probably not be suitable for foreign students (i.e., not Anglo-Saxon and in particular French). I was wrong, *mea culpa*. Thank you for your efforts concerning the use of alternative forms of learning."

All 15 instructors involved in this study volunteered to introduce ILS into their classroom. Half of the instructors indicated that they significantly changed their courses, while the others simply adjusted their course to create time for TPS questions. The instructors were also asked whether they were willing to continue using interactive learning in the future years and all the instructors agreed that they would, which gives significant momentum to the Physics Department to continue promoting the use of ILS at UPMC. This last result is particularly significant given the finding that in the U.S., a third of instructors who try research-based instructional strategies (RBIS), including the ILS used in our study, discontinue use after trying them at least once [25]. Though we have no concrete evidence to explain our high (100%) continuation rate, studies of change strategies in higher education show that successful strategies incorporate support during implementation and feedback [34]; thus, we suggest that the support and feedback we provided, through the initial intensive instructor training, freely available in-semester support, and an end-of-semester debriefing session, may have played an important role in promoting continuation of the use of ILS among the instructors at UPMC.

## VII. CONCLUSIONS

We have conducted a study of interactive learning in two large introductory physics classes in a major French university, a first-year, second-semester mechanics class, and a second-year E&M class. In both classes, some instructors utilized ILS, while others continued teaching in a more traditional style (primarily lecture), the latter constituting a natural control group. We provided introductory training to the instructors implementing ILS in their classes via two training workshops, and supported those instructors throughout the semester by conducting classroom visits, at their request, or consulting with instructors who asked for help or feedback. We administered a research-validated concept inventory in each class (FCI for mechanics and CSEM for E&M), as well as collecting final exam scores. We also administered demographics and attitude surveys to the students in both classes, and an attitude survey to the instructors utilizing ILS in their class.

Our two main conclusions are:

1. Interactive learning had a positive effect on student learning gains in two distinct large introductory physics classes, by two distinct measures: performance on research-validated concept inventories; and performance on final exams, both conceptual and traditional problem-solving questions. The presence or level of interactivity in the classroom had among the largest, if not the largest, predictive strength for student learning among the factors we considered in four different multivariate models, including parents' education, GPA, and hours studied per week.

2. Both students and instructors had very positive impressions of the use of ILS in their class. Both groups indicated that they believed that ILS improved student learning and student assiduousness in class, and the students in classes implementing ILS indicated a higher interest in physics compared to those in traditional classes.

Overall the positive outcomes of this study in an educational setting very different from that found in most U.S. colleges and universities is encouraging, supporting the contention that ILS are designed to address how *people* learn, whether in France or the U.S. While it would be an overstatement to say that our study proves that ILS will work in all educational settings around the world, it certainly shows that cultural influences or differences in educational systems need not be a barrier to the effective implementation of interactive learning strategies in university physics classrooms outside the U.S.

# ACKNOWLEDGEMENTS

The authors want to begin by thanking the instructors and students at UPMC who participated in this study. Without their participation, none of this work would have been possible. We wish to particularly thank the instructors, who devoted a significant amount of time and effort challenging themselves to try a novel and time-consuming pedagogical innovation, in pursuit of improving their students' learning. We also wish to thank the chair of the Physics Department, Dr. Patrick Boissé for his support for this project; Dr. Catherine Schwob, who installed clickers in the *Amphi* used by the mechanics classes; Dr. Edouard Kierlick, who purchased the clickers used in the E&M classes, and who encouraged instructor use of ILS; Dr. Yves Noël who helped administer the online CSEM test; Dr. Nicole Poteaux who helped in designing the demographics questions, and *Turning Technologies,* who lent clickers for the study presented here.

We wish to thank Lillian McDermott, Peter Shaffer, Paula Heron, and the members of the University of Washington Physics Education Group for their support and permission to use their *Tutorials in Introductory Physics*, and Dr. Rachel Scherr for her help providing videos from the *Video Resource for Professional Development of University Physics Educators* for use in the instructor-training workshops. We also thank Drs. Scherr and Heron, Dr. Steven Pollock, as well as the referees, for comments that greatly improved the paper.

Finally, Dr. Rudolph wishes to thank Dr. Anne-Laure Melchior for facilitating his visit to France, and to thank the faculty, staff, and administration of the Université Pierre et Marie Curie for their hospitality during his stay in Paris. He particularly wants to thank the chair of the Physics Department, Dr. Patrick Boissé for support for his visit to UPMC.

This material is based in part upon work supported by the National Science Foundation under Grant No. AST-0847170. Any opinions, findings, and conclusions or recommendations expressed in this material are those of the authors and do not necessarily reflect the views of the National Science Foundation.

---